\documentclass[a4paper,11pt]{article}

\usepackage{amsmath,amssymb}     
\usepackage{color}
\usepackage{graphicx}
\usepackage{subfigure}
\usepackage{cite}                
\usepackage{hyperref}            
\usepackage{multirow,makecell}   

\numberwithin{equation}{section}   

\def \be {\begin{equation}}
\def \ee {\end{equation}}
\def \ba {\begin{array}}
\def \ea {\end{array}}
\def \bea{\begin{eqnarray}}
\def \eea{\end{eqnarray}}
\def \nn {\nonumber}

\def \a {\alpha}
\def \b {\beta}
\def \g {\gamma}

\def \d {\delta}

\def \e {\epsilon}

\def \m {\mu}
\def \n {\nu}

\def \l {\lambda}

\def \s {\sigma}

\def \r {\rho}

\def \th {\theta}
\def \vth {\vartheta}

\def \t {\tau}
\def \z {\zeta}

\def \mA {\mathcal A}

\def \mD {\mathcal D}

\def \mN {\mathcal N}

\def \mP {\mathcal P}

\def \p {\partial}
\def \f {\frac}

\def \mc {\mathcal}

\def \lt {\left}
\def \rt {\right}

\def \sr {\sqrt}
\def \td {\tilde}

\def \dd {\mathrm{d}}

\def \tr {\textrm{tr}}

\def \diag {{\textrm{diag}}}

\def \ph {\phantom}

\begin{document}

\title{\textbf{BPS Wilson loops in Minkowski spacetime and Euclidean space}}
\author{
Hao Ouyang\footnote{ouyangh@ihep.ac.cn},
Jun-Bao Wu\footnote{wujb@ihep.ac.cn}~
and
Jia-ju Zhang\footnote{jjzhang@ihep.ac.cn}
}
\date{}

\maketitle

\vspace{-10mm}

\begin{center}
{\it
Theoretical Physics Division, Institute of High Energy Physics, Chinese Academy of Sciences,\\
19B Yuquan Rd, Beijing 100049, P.~R.~China\\\vspace{1mm}
Theoretical Physics Center for Science Facilities, Chinese Academy of Sciences,\\19B Yuquan Rd, Beijing 100049, P.~R.~China
}
\vspace{10mm}
\end{center}

\begin{abstract}

  We give evidence that spacelike BPS Wilson loops do not exist in Minkowski spacetime. We show that spacelike Wilson loops in Minkowski spacetime cannot preserve any supersymmetries, in $d = 4$ $\mathcal N = 4$ super Yang-Mills theory, $d = 3$ $\mathcal N = 2$ super Chern-Simons-matter theory, and $d = 3$ $\mathcal N = 6$ Aharony-Bergman-Jafferis-Maldacena theory. We not only show this using infinite straight lines and circles as examples, but also we give proofs for general curves. We attribute this to the conflicts of reality conditions of the spinors. However, spacelike Wilson loops do exist in Euclidean space. There are both BPS Wilson loops along infinite straight lines and circular BPS Wilson loops. This is because the reality conditions of the spinors are lost after Wick rotation. The result is reasonable in view of the AdS/CFT correspondence.

\end{abstract}


\baselineskip 18pt
\thispagestyle{empty}

\newpage

\tableofcontents

\section{Introduction}

Supersymmetric Wilson loops play an important role in the AdS/CFT correspondence, as they are dual to fundamental strings or membranes in the bulk string/M theory.
In the AdS$_5$/CFT$_4$ correspondence, type IIB string theory in AdS$_5\times$S$^5$ spacetime is dual to $d=4$ $\mN=4$ super Yang-Mills (SYM) theory\cite{Maldacena:1997re,Gubser:1998bc,Witten:1998qj}. There are 1/2 BPS Wilson loops in $d=4$ $\mN=4$ SYM theory, and they are supposed to be dual to  worldsheets of fundamental strings in type IIB string theory\cite{Maldacena:1998im,Rey:1998ik}.
Similarly, in the AdS$_4$/CFT$_3$ correspondence, M-theory in AdS$_4\times$S$^7$/Z$_k$ spacetime, or type IIA string theory in AdS$_4\times$CP$^3$ spacetime, is dual to $d=3$ $\mN=6$ super Chern-Simons-matter (SCSM) theory with levels $(k,-k)$ and gauge group $U(N)\times U(N)$, which is known as Aharony-Bergman-Jafferis-Maldacena (ABJM) theory\cite{Aharony:2008ug}.
In ABJM theory there are 1/6 BPS Wilson loops \cite{Drukker:2008zx,Chen:2008bp,Rey:2008bh}, which are dual to smeared fundamental strings in type IIA string theory\cite{Drukker:2008zx}. This kind of 1/6 BPS Wilson loops in ABJM theory are closely related to the 1/2 BPS Wilson loops in $\mN=2$ SCSM theory proposed in \cite{Gaiotto:2007qi}. They are constructed solely by bosonic fields, and we will call them Gaiotto-Yin (GY) type Wilson loops.
There are also 1/2 BPS Wilson loops in ABJM theory that are dual to the simplest fundamental strings in type IIA string theory \cite{Drukker:2009hy}. There are fermionic fields in such loops, and we will call them Drukker-Trancanelli (DT) type Wilson loops.

In this paper we revisit the Wilson loops in the AdS/CFT correspondence. We find that in  Minkowski spacetime, there is no BPS spacelike Wilson loop.
This arises from conflicts of reality conditions of spinors.
We first show this using infinite straight lines and circles as examples. Then we give proofs that in Minkowski spacetime BPS Wilson loops along general curves are necessarily timelike or null. The proofs are general for Wilson loops in $d=4$ $\mN=4$ SYM theory, $d=3$ $\mN=2$ SCSM theory, and GY type and DT type Wilson loops in ABJM theory. However, for the Euclidean version of the AdS/CFT correspondence, there is another story. The reality conditions of the spinors are lost in the Wick rotation, and the conflicts in Minkowski spacetime disappear in Euclidean space.
Then we can have BPS Wilson loops along spacelike curves.


The rest of this paper is organized as follows.
In Section~\ref{s2} we discuss Wilson loops along infinite straight lines and circles in Minkowski spacetime. We show that there are timelike and null BPS Wilson loops, but there are no spacelike ones.
In Section~\ref{s3} we present the case in Euclidean space. We show that spacelike BPS Wilson loops along infinite straight lines and circles are allowed in Euclidean space.
In Section~\ref{general}, we give general proofs that BPS Wilson loops in Minkowski spacetime are necessarily timelike or null, but not spacelike.
We end with conclusion and discussion in Section~\ref{s4}. In Appendix~\ref{sa} we review the definition of Majorana spinors in various dimensions. In Appendix~\ref{sb} we discuss the consistency of constraints for Majorana spinors. In Appendix~\ref{sc} and \ref{sd}, there are spinor conventions in $d=3$ Minkowski spacetime and Euclidean space, respectively.

\section{Straight lines and circles in Minkowski spacetime}\label{s2}

In this section we consider the BPS Wilson loops along infinite straight lines and circles in several supersymmetric conformal field theories in Minkowski spacetime. Thanks to the conformal symmetry, it is enough to consider Wilson loops along the timelike infinite straight line $x^\mu(\tau) = \tau \delta^\mu_0$, null line $x^\mu(\tau) = \tau(\delta^\mu_0+\delta^\mu_1)$ and spacelike line $x^\mu(\tau) = \tau\delta^\mu_1$. There are conformal transformations that map infinite straight lines to circles.

\subsection{\emph{d}=4 $\bf\mN$=4 SYM theory}

We use the $d=10$ $\mN=1$ SYM theory formalism. The supersymmetry (SUSY) transformation of the bosonic fields is
\be \label{susyde4ne4}
\d_\e A_\m=\bar\l\g_\m\e, ~~~ \d_\e \phi_I=\bar\l\g_I\e.
\ee
Here $A_\m$ with $\m=0,1,2,3$ is a four-dimensional vector, $\phi_I$ with $I=4,\cdots,9$ are four-dimensional scalars, $\l$ is the fermionic field, and $\bar\l=\l^\dagger\g^0$. The symbols $\g_\m$ and $\g_I$ are ten-dimensional gamma matrices, and $\e$ is the SUSY transformation parameter. Also $\l$ and $\e$ are ten-dimensional Weyl-Majorana spinors.

One may define the Wilson loop\cite{Maldacena:1998im,Rey:1998ik}
\be \label{wlde4ne4}
W=\mP \exp \lt[ {ig\int d\t \lt( A_\m \dot x^\m+\phi_I y^I |\dot x| \rt)} \rt],
\ee
with $\mP$ denoting path-ordering operation.

For a timelike infinite straight line one may choose
\be \label{yi1}
x^\m=\t \d^\m_0, ~~~ y^I=\d^I_4.
\ee
Then, $\d_\e W=0$ leads to
\be
\g_{04}\e=\e.
\ee
The choice of $y^I$ in (\ref{yi1}) is consistent with the fact that $(\g_{04})^2=1$.
From analyses in Appendix~\ref{sb}, $\g_{04}$ is a consistent constraint matrix for a Majorana spinor in $d=10$ Minkowski spacetime. Thus it is a timelike 1/2 BPS Wilson loop.

Also one may choose a null infinite straight line $x^\m=\t (\d^\m_0+\d^\m_1)$. In this case $|\dot x|=0$, the $y^I$ term vanishes, and the Wilson loop (\ref{wlde4ne4}) becomes
\be
W=\mP \exp \lt( {ig\int d\t A_\m \dot x^\m } \rt).
\ee
Now $\d_\e W=0$ leads to
\be
\g_{01}\e=\e.
\ee
This is also legal, and it is a null 1/2 BPS Wilson loop.

To get a spacelike Wilson loop we choose
\be \label{z1}
x^\m=\t \d^\m_1, ~~~ y^I=i \d^I_4.
\ee
The constraint from $\d_\e W=0$ now becomes
\be \label{z2}
\g_{14}\e=i\e.
\ee
The choice of $y^I$ in (\ref{z1}) is consistent with the fact that $(\g_{14})^2=-1$.
From Appendix~\ref{sb} it is illegal because it contradicts the reality conditions of Majorana spinors in $d=10$ Minkowski spacetime. So there is no spacelike 1/2 BPS Wilson loop along an infinite straight line in $\mN=4$ SYM theory in $d=4$ Minkowski spacetime.
Since the $d=4$ $\mN=4$ SYM theory is a superconformal theory and an infinite straight line can be mapped to a circle by an appropriate conformal transformation, there is no spacelike 1/2 BPS Wilson loop along a circle either.\footnote{Similar consideration appeared in \cite{Munkler:2013}, while clear conclusion was lacked there.}

\subsection{\emph{d}=3 $\bf\mN$=2 SCSM theory}

We consider the $d=3$ $\mN=2$ SCSM theory that has a gauge field $A_\m$, a complex scalar $\phi$ and a Dirac spinor $\psi$. It is also a superconformal theory, and one can write a general superconformal transformation as
\bea \label{susyne2}
&& \d_\chi A_\m=-\f{2\pi}{k} (\phi\bar\psi\g_\m\chi+\bar\chi\g_\m\psi\bar\phi), \nn\\
&& \d_\chi \phi=i\bar\chi\psi, ~~~ \d_\chi \bar\phi=i\bar\psi\chi,
\eea
with $\chi=\th+x^\m\g_\m\vth$, $\bar\chi=\bar\th-\bar\vth x^\m\g_\m$, and $\th$, $\vth$ being constant Dirac spinors. The $\th$, $\bar\th$ terms are the Poincar\'e SUSY transformations, and $\vth$, $\bar\vth$ terms are conformal SUSY transformations.
The transformations $\d_\chi\psi$ and $\d_\chi\bar\psi$ will not be used, and so we will not bother to write them out. Note that $A_\m=A_\m^\dagger$, $\bar\phi=\phi^\dagger$, and the SUSY transformation preserves these relations $\d_\chi A_\m=\d_\chi A_\m^\dagger$, $\d_\chi\bar\phi=\d_\chi\phi^\dagger$.

The 1/2 BPS Wilson loop in $\mN=2$ SCSM theory was found in \cite{Gaiotto:2007qi}, and we first give a short review in this subsection.
One can define the 1/2 BPS Wilson loop as
\bea \label{w12ne2}
&& W=\mP \exp\lt( -i\int d\t\mA(\t) \rt) ,\nn\\
&& \mA=A_\m \dot x^\m+\f{2\pi}{k} m\phi\bar\phi |\dot x|.
\eea

One can consider the timelike straight line $x^\m=\t \d^\m_0$.
For the Poincar\'e SUSY transformation invariance of $W$, i.e. $\d_\th\mA=0$, one gets
\be
\g_0\th=i m\th, ~~~
\bar\th\g_0=i m\bar\th.
\ee
The second equation is just
\be\g_0\bar\th=-i m\bar\th. \ee
Since the eigenvalues of $\g_0$ are $\pm i$, for $\th\neq0$ one can only have $m=\pm1$. One can choose $ m=1$ without loss of generality.
Thus one gets
\be
\g_0\th=i\th, ~~~
\g_0\bar\th=-i\bar\th.
\ee
They are compatible, since they are just the complex conjugates of each other.
It is similar for the conformal SUSY transformation.
So one gets a 1/2 BPS Wilson loop along a timelike infinite straight line.

We may choose a null infinite straight line $x^\m=\t (\d^\m_0+\d^\m_1)$. In this case $|\dot x|=0$, the $m$ term vanishes, and (\ref{w12ne2}) becomes
\be
W=\mP \exp \lt( {-i\int d\t A_\m \dot x^\m } \rt),
\ee
and now the Poincar\'e SUSY transformation $\d_\th W=0$ leads to
\be
\g_2 \th=-\th, ~~~ \g_2 \bar\th=-\bar\th,
\ee
which are compatible. It is similar for the conformal SUSY. We have a 1/2 BPS Wilson loop along  a null infinite straight line.

Then we consider the Wilson loop (\ref{w12ne2}) along a spacelike infinite straight line $x^\m=\t \d^\m_1$.
For the Poincar\'e SUSY transformation invariance of $W$, we get
\be
\g_1\th=i m\th, ~~~
\g_1\bar\th=-i m\bar\th.
\ee
The eigenvalues of $\g_1$ are $\pm 1$, and without loss of generality we choose $ m=-i$.
Thus we obtain
\bea \label{z3}
\g_1\th=\th, ~~~
\g_1\bar\th=-\bar\th.
\eea
Note that they are not compatible. It is similar for conformal SUSY. So we conclude that there is no 1/2 BPS Wilson loop along a spacelike infinite straight line in $\mN=2$ SCSM theory. There is conformal transformation that turns a spacelike infinite straight line into a circle, and so there is no 1/2 BPS Wilson loop along a spacelike circle either.

\subsection{ABJM theory}

The ABJM theory is an $\mN=6$ SCSM theory, and it was constructed in \cite{Aharony:2008ug}.
ABJM theory has $U(N)\times U(N)$ gauge symmetry, and the gauge fields are $A_\m$ and $\hat A_\m$ respectively.
The complex scalar $\phi_I$ and Dirac spinor $\psi_I$ are in $(N,\bar N)$ bifundamental representation,
and so $\bar\phi^I=\phi_I^\dagger$ and $\bar\psi_I=(\psi^I)^\dagger$ are in the $(\bar N, N)$ representation.
We have used $I,J,K,L,\cdots=1,2,3,4$ as indices of the $SU(4)$ R-symmetry.
A general superconformal transformation of ABJM theory is \cite{Gaiotto:2008cg,Hosomichi:2008jb,Terashima:2008sy,Bandres:2008ry}
\bea \label{abjmsusy}
&& \d_\chi A_\m=\f{2\pi}{k} \lt( \phi_I\bar\psi_J\g_\m\chi^{IJ} +\bar\chi_{IJ}\g_\m\psi^J\bar\phi^I \rt), \nn\\
&& \d_\chi\hat A_\m=\f{2\pi}{k} \lt( \bar\psi_J\g_\m\phi_I\chi^{IJ}+\bar\chi_{IJ}\bar\phi^I\g_\m\psi^J \rt), \nn\\
&& \d_\chi\phi_I=i\bar\chi_{IJ}\psi^J, ~~~
   \d_\chi\bar\phi^I=i\bar\psi_J\chi^{IJ},\\
&& \d_\chi\psi^I=\g^\m\chi^{IJ}D_\m\phi_J +\vth^{IJ}\phi_J
            -\f{2\pi}{k}\chi^{IJ} \lt( \phi_J\bar\phi^K\phi_K-\phi_K\bar\phi^K\phi_J \rt)
            -\f{4\pi}{k}\chi^{KL}\phi_K\bar\phi^I\phi_L, \nn\\
&& \d_\chi\bar\psi_I=-\bar\chi_{IJ}\g^\m D_\m\bar\phi^J +\bar\vth_{IJ}\bar\phi^J
                +\f{2\pi}{k}\bar\chi_{IJ} \lt( \bar\phi^J\phi_K\bar\phi^K-\bar\phi^K\phi_K\bar\phi^J \rt)
                +\f{4\pi}{k}\bar\chi_{KL}\bar\phi^K\phi_I\bar\phi^L, \nn
\eea
with $\chi^{IJ}=\th^{IJ}+x^\m\g_\m\vth^{IJ}$ and $\bar\chi_{IJ}=\bar\th_{IJ}-\bar \vth_{IJ}x^\m\g_\m$.
The definitions of covariant derivatives are
\bea
&& D_\m\phi_J =\p_\m \phi_J +i A_\m \phi_J -i \phi_J \hat A_\m ,\nn\\
&& D_\m\bar\phi^J=\p_\m\bar\phi^J +i \hat A_\m \bar\phi^J -i \bar\phi^J  A_\m.
\eea
Also $\th^{IJ}$, $\bar\th_{IJ}$ and $\vth^{IJ}$, $\bar\vth_{IJ}$ are Dirac spinors with constraints
\bea \label{thij}
&& \th^{IJ}=-\th^{JI}, ~~~ (\th^{IJ})^*=\bar \th_{IJ}, ~~~ \bar\th_{IJ}=\f{1}{2}\e_{IJKL}\th^{KL}, \nn\\
&& \vth^{IJ}=-\vth^{JI}, ~~~ (\vth^{IJ})^*=\bar \vth_{IJ}, ~~~ \bar\vth_{IJ}=\f{1}{2}\e_{IJKL}\vth^{KL}.
\eea
The $\e_{IJKL}$ symbole is totally antisymmetric with $\e_{1234}=1$.
Like the $\mN=2$ SCSM theory, the $\th$, $\bar\th$ terms is Poincar\'e SUSY transformation, and $\vth$, $\bar\vth$ terms is conformal SUSY transformation. Note that we have $\d_\chi A_\m=\d_\chi A_\m^\dagger$, $\d_\chi \hat A_\m=\d_\chi \hat A_\m^\dagger$, $\d_\chi\bar\phi^I=\d_\chi\phi_I^\dagger$, and $\d_\chi\bar\psi^I=\d_\chi\psi_I^\dagger$.

\subsubsection{1/6 BPS Wilson loop}

The 1/6 BPS Wilson loop along an infinite straight line is the GY type Wilson loop.
It was constructed in \cite{Drukker:2008zx,Chen:2008bp,Rey:2008bh},
and there was a careful analysis of reality conditions for the spinors $\th_{IJ}$ and $\bar \th^{IJ}$ in \cite{Rey:2008bh}.
It is closely related to the Wilson loop constructed in \cite{Gaiotto:2007qi}, which we have reviewed in the previous subsection.
The Wilson loop takes the form
\bea \label{w16a}
&& W=\mP \exp \lt( -i\int d\t\mA(\t) \rt),\nn\\
&& \mA=A_\m \dot x^\m +\f{2\pi}{k} M^I_{\ph{I}J} \phi_I\bar\phi^J |\dot x|.
\eea

For a timelike straight line $x^\m=\t \d^\m_0$, it can be shown that for
\bea
&& M^I_{\ph{I}J}=\diag( -1,-1,1,1 ),  \nn\\
&& \g_0\th^{12}=i\th^{12}, ~~~ \g_0\th^{34}=-i\th^{34},  \nn\\
&& \th^{13}=\th^{14}=\th^{23}=\th^{24}=0,\\
&& \g_0\vth^{12}=i\vth^{12}, ~~~
   \g_0\vth^{34}=-i\vth^{34}, \nn\\
&& \vth^{13}=\vth^{14}=\vth^{23}=\vth^{24}=0, \nn
\eea
and here we have $\d_\chi\mA=0$, and so $\d_\chi W=0$. Thus Wilson loop (\ref{w16a}) along a timelike infinite straight line is 1/6 BPS.

Similarly, there can be null BPS Wilson loop along $x^\m=\t (\d^\m_0+\d^\m_1)$, and the preserved SUSY is
\be
\g_2 \th^{IJ}=-\th^{IJ}, ~~~ \g_2 \vth^{IJ}=-\vth^{IJ}.
\ee
Note that now we get $M^I_{\ph{I}J}|\dot x|=0$, and so $\mA=A_\m \dot x^\m$.
Thus we have a null BPS Wilson loop. It is not 1/6 BPS, but 1/2 BPS.

On the other hand if we want a spacelike Wilson loop $x^\m=\t \d^\m_1$, for Poincar\'e SUSY transformation we have
\be
\d_\th\mA=\f{2\pi}{k} \lt[ \phi_K\bar\psi_J \lt( \d^K_I\g_1+i M^K_{\ph K I} \rt) \th^{IJ}
                       +\bar\th_{IJ} \lt( \d^I_K\g_1+i M^I_{\ph I K} \rt) \psi^J\bar\phi^K \rt].
\ee
Thus for $\d_\th\mc A=0$ we have
\be
\g_1 \th^{IJ}=-iM^I_{\ph I K}\th^{KJ}, ~~~
\g_1 \bar\th_{IJ}=i M^K_{\ph K I}\bar\th_{KJ}.
\ee
In the basis of diagonalized $M^I_{\ph I J}= m_I \d^I_J$, we get
\be \label{e1}
\g_1 \th^{IJ}=-i m_I \th^{IJ}, ~~~
\g_1 \bar\th_{IJ}=i  m_I \bar\th_{IJ},
\ee
with no index summations on the right hand sides of the two equations.
Note that the eigenvalues of $\g_1$ are $\pm 1$.
Without loss of generality we may suppose $\th^{12}\neq 0$ and get
\be
\g_1 \th^{12}= \th^{12}.
\ee
Using (\ref{thij}) we get $(\th^{12})^*=\th^{34}$. Then we obtain
\be
\g_1\th^{34}=\th^{34}.
\ee
This means that all $ m_I=i$. Then (\ref{e1}) become
\be 
\g_1 \th^{IJ}=\th^{IJ}, ~~~
\g_1 \bar\th_{IJ}=-\bar\th_{IJ},
\ee
which are not consistent. It is similar for conformal SUSY transformation.
So there is no 1/6 BPS spacelike Wilson loop along an infinite straight line.
There is no 1/6 BPS spacelike Wilson loop along a circle either.

Similarly one can construct the 1/6 BPS Wilson loop along $x^\m=\t \d^\m_0$
\bea \label{w16ha}
&& \hat W=\mP \exp \lt( -i\int d\t \hat\mA(\t) \rt),\nn\\
&& \hat\mA=\hat A_\m \dot x^\m +\f{2\pi}{k} N_I^{\ph{I}J} \bar\phi^I\phi_J |\dot x|, \nn\\
&& N_I^{\ph{I}J}=\diag( -1,-1,1,1 ).
\eea
It is a timelike BPS Wilson loop similar to (\ref{w16a}), and the only difference is that it involves $\hat A_\m$ instead of $A_\m$.
It preserves the same SUSY as (\ref{w16a}).
For a null infinite straight line one has $\hat\mA=\hat A_\m \dot x^\m$, and it is 1/2 BPS.
But still no spacelike 1/6 BPS Wilson loop along a straight line or a circle is allowed.

\subsubsection{1/2 BPS Wilson loop}

The 1/2 BPS Wilson loop  besides the null case was constructed in \cite{Drukker:2009hy}.
Such construction of BPS Wilson loops was explained elegantly via the Brout-Englert-Higgs mechanism in \cite{Lee:2010hk}.

One considers the Wilson loop along the timelike infinite straight line $x^\m=\t \d^\m_0$
\be \label{w12abjm}
W=\mP \exp \lt( -i\int d\t L(\t) \rt),
\ee
where $L$ is a supermatrix
\be
L=\lt( \ba{cc} \mA &\bar f_1 \\ f_2 & \hat\mA \ea \rt).
\ee
Here we defined
\bea
&& \mA=A_\m\dot x^\m+\f{2\pi}{k} M^I_{\ph IJ} \phi_I\bar\phi^J |\dot x|, \nn\\
&& \hat\mA=\hat A_\m\dot x^\m+\f{2\pi}{k} N_I^{\ph IJ}\bar\phi^I \phi_J |\dot x|, \\
&& \bar f_1=\sr{\f{2\pi}{k}}\bar\zeta_I\psi^I |\dot x|, ~~~
   f_2=\sr{\f{2\pi}{k}}\bar\psi_I\eta^I |\dot x|.  \nn
\eea
Note that $\bar\zeta_I$ and $\eta^I$ are Grassmann even, and so $\bar f_1$ and $f_2$ are Grassmann odd.
To make $W$ SUSY invariant\footnote{We assume that all fields tend  to zero as $\t \to \pm \infty$.} $\d_\chi L=0$ is not necessary, and it is enough to require that\footnote{Notice that for the Wilson loops in the previous subsections, such relaxation does not give anything new since the SUSY transformation of gauge fields and scalar fields does not involve any derivatives. }\cite{Lee:2010hk}
\be
\d_\chi L=\p_\t G+i[L,G],
\ee
for some Grassmann odd matrix
\be
G= \lt( \ba{cc} & \bar g_1 \\ g_2 &  \ea \rt).
\ee
Concretely, one needs
\bea \label{e2}
&& \d_\chi \mA=i(\bar f_1 g_2-\bar g_1 f_2),  \nn\\
&& \d_\chi \hat\mA=i(f_2 \bar g_1 - g_2 \bar f_1),  \\
&& \d_\chi \bar f_1 = \p_\t \bar g_1+i\mA \bar g_1-i\bar g_1\hat\mA,  \nn\\
&& \d_\chi f_2 = \p_\t g_2+i\hat\mA g_2-ig_2\mA.  \nn
\eea

As in \cite{Drukker:2009hy}, one can use symmetry to guide the search for a 1/2 BPS Wilson loop.
One can break the $SU(4)$ R-symmetry to $U(1)\times SU(3)$ by writing $I=(1,i)$ with $i=2,3,4$.%
\footnote{The $SU(3)$ R-symmetry invariance is necessary if one requires that the Wilson loop has simple M2 brane dual in M theory. If the M2 brane does not stretch in the compactified space S$^7$/Z$_k$ except M-theory circle, it would be global $SU(3)$ R-symmetry. Otherwise the $SU(3)$ R-symmetry would be only local. In Subsection~\ref{s4.3} we will consider more general Wilson loops and do not require that the Wilson loops are invariant under any subgroups of the $SU(4)$ R-symmetry.}
For general $\bar\zeta_I$, $\eta^I$, $M^I_{\ph{I}J}$ and $N_I^{\ph{I}J}$, the $SU(4)$ R-symmetry will be broken totally. One wishes to get a BPS Wilson loop with the global $SU(3)$ subgroup intact, and so one can choose
\bea
&& \bar\zeta_I=\bar\zeta\d^1_I, ~~~ \eta^I=\eta\d^I_1, \nn\\
&& M^I_{\ph{I}J}=\diag ( m_1, m_2, m_2, m_2 ), \\
&& N_I^{\ph{I}J}=\diag ( n_1, n_2, n_2, n_2 ).\nn
\eea
One can suppose the $SU(3)$ invariant constraint
\be
\g_0\th^{1i}=i\th^{1i},
\ee
and then from (\ref{thij}) one obtains
\be
\g_0\th^{ij}=-i\th^{ij}, ~~~\bar \th_{1i}\g_0=i\bar\th_{1i}, ~~~ \bar \th_{ij}\g_0=-i\bar\th_{ij},
\ee
Since $\psi^i$ and $\bar\psi_i$ do not appear in $\bar f_1$ and $f_2$, to satisfy (\ref{e2}) they must not appear in $\d_\th\mA$ or $\d_\th\hat\mA$ either. So one has to choose $ m_1=n_1=-1$ and $ m_2=n_2=1$, and then for Poincar\'e SUSY transformation one can get
\bea
&& \d_\th\mA=-\f{4\pi i}{k} \lt( \phi_i\bar\psi_1\th^{1i}+\bar\th_{1i}\psi^1\bar\phi^i \rt), \nn\\
&& \d_\th\hat\mA=-\f{4\pi i}{k} \lt(\bar\psi_1\th^{1i}\phi_i+\bar\phi^i\bar\th_{1i}\psi^1 \rt).
\eea
In order that $\d_\th \bar f_1$  and $\d_\th f_2$ satisfy the form of (\ref{e2}), one must choose
\be
\g_0 \eta=i\eta, ~~~ \bar\zeta \g_0=i\bar\zeta.
\ee
Then one gets
\bea
&& \d_\th \bar f_1=-i\sr{\f{2\pi}{k}}\bar\zeta\th^{1i} \mD_0\phi_i, ~~~
   \bar g_1=-i\sr{\f{2\pi}{k}}\bar\zeta\th^{1i} \phi_i, \nn\\
&& \d_\th f_2=i\sr{\f{2\pi}{k}}\bar\th_{1i}\eta \mD_0\bar\phi^i, ~~~
   g_2=i\sr{\f{2\pi}{k}}\bar\th_{1i}\eta \bar\phi^i .
\eea
One can show that, given
\be \label{etabareta}
\eta \bar\zeta=-i-\g_{0},
\ee
the equations (\ref{e2}) are satisfied.
It is similar for conformal SUSY transformation.
Thus the Wilson loop (\ref{w12abjm}) along a timelike infinite straight line is indeed 1/2 BPS.

For a null infinite straight line $|\dot x|=0$, Wilson loop (\ref{w12abjm}) becomes trivially the same as discussed before, and it is 1/2 BPS.

If we want to repeat the above calculation with the spacelike straight line $x^\m=\t \d^\m_1$ in Minkowski spacetime, we will run into dilemma. Now we suppose
\be
\g_1\th^{1i}=\th^{1i},
\ee
and then we get
\be
\g_1\th^{ij}=\th^{ij}, ~~~\bar \th_{1i}\g_1=-\bar\th_{1i}, ~~~ \bar \th_{ij}\g_1=-\bar\th_{ij}.
\ee
Then we cannot choose $m_{1,2}$, $n_{1,2}$ to make $\psi^i$ and $\bar\psi_i$ do not appear in $\d_\th\mA$ and $\d_\th\hat\mA$. So we conclude that there is no 1/2 BPS Wilson loop with global $SU(3)$ R-symmetry along a spacelike straight line. There is no 1/2 BPS Wilson loop with global $SU(3)$ R-symmetry along a spacelike circle either.

\section{Straight lines and circles in Euclidean space}\label{s3}

In the previous section we failed in search of spacelike BPS Wilson loops in Minkowski spacetime because of contradictions of reality conditions for spinors. However, the reality conditions for spinors in Minkowski spacetime disappear if we go to Euclidean space by a Wick rotation \cite{Osterwalder:1973kn,Nicolai:1978vc}.
In this section we show explicitly that BPS Wilson loops along spacelike infinite straight lines and circles exist in Euclidean space.

\subsection{\emph{d}=4 $\bf\mN$=4 SYM theory}

For the Euclidean $d=4$ $\mN=4$ SYM theory the SUSY transformation is formally identical to (\ref{susyde4ne4}), but now $\l$ and $\e$ are no longer Majorana spinors. 
Explicitly, we have
\be
\e \neq \e_c
\ee
with charge conjugate being defined as (\ref{thc}). The Wilson loop is defined formally the same as (\ref{wlde4ne4}), and for the infinite straight line (\ref{z1}), $\d_\e W=0$ still leads to (\ref{z2}). It now becomes legal, since $\e$ is no longer a Majorana spinor.
There exist conformal transformations that take an infinite straight line to a circle, and so there are circular BPS Wilson loops in the Euclidean $d=4$ $\bf\mN=4$ SYM theory too.
The spacelike BPS Wilson loops are just the ones that were studied in \cite{Maldacena:1998im,Rey:1998ik,Berenstein:1998ij,Drukker:1999zq}.

\subsection{\emph{d}=3 $\bf\mN$=2 SCSM theory}

The Euclidean $d=3$ $\mN=2$ SCSM theory has formally identical SUSY transformation as (\ref{susyne2}), but now $\bar\psi$ is not related to $\psi$, $\bar\chi$ is not related to $\chi$, $\bar\th$ is not related to $\th$, and $\bar\vth$ is not related to $\vth$. Although we have $A_\m=A_\m^\dagger$, $\bar\phi=\phi^\dagger$, but we do not have $\d_\chi A_\m=\d_\chi A_\m^\dagger$ or $\d_\chi\bar\phi=\d_\chi\phi^\dagger$. Formally we can define a Wilson loop as the Minkowski case (\ref{w12ne2}), and now for a straight line (\ref{z3}) is legal, since $\bar\th$ and $\th$ are not related. So there are 1/2 BPS Wilson loops along infinite straight lines, as well as circular 1/2 BPS Wilson loops in Euclidean $d=3$ $\mN=2$ SCSM theory. For the Wilson loop along the line $x^\m=\t \d^\m_1$, the preserved SUSY is
\be
\g_1 \th=\th, ~~~ \bar\th\g_1=\bar\th, ~~~ \g_1 \vth=\vth, ~~~ \bar\vth\g_1=\bar\vth.
\ee
For the circular Wilson loop $x^\m=(\cos\t,\sin\t,0)$, the preserved SUSY is
\be
\vth=i\g_3\th, ~~~ \bar \vth=i\bar\th\g_3.
\ee

\subsection{ABJM theory}

For the Euclidean ABJM theory, it is similar to the above two cases.  The  superconformal transformation is formally the same as (\ref{abjmsusy}), with $\chi^{IJ}=\th^{IJ}+x^\m\g_\m\vth^{IJ}$ and $\bar\chi_{IJ}=\bar\th_{IJ}-\bar \vth_{IJ}x^\m\g_\m$.
But now (\ref{thij}) becomes
\bea \label{z4}
&& \th^{IJ}=-\th^{JI}, ~~~ \bar\th_{IJ}=\f{1}{2}\e_{IJKL}\th^{KL}, \nn\\
&& \vth^{IJ}=-\vth^{JI}, ~~~ \bar\vth_{IJ}=\f{1}{2}\e_{IJKL}\vth^{KL}.
\eea
Note that the twelve spinors $\th^{[IJ]}$, $\vth^{[IJ]}$ with $I,J=1,2,3,4$ are independent Dirac spinors.

For the Wilson loop (\ref{w16a}) of infinite straight line $x^\m=\t \d^\m_1$\footnote{Notice that as mentioned in Appendix~\ref{sd}, the components of the  coordinates $x^\mu$ are denoted as $(x^1, x^2, x^3)$ in $d=3$ Euclidean space.},
we can simultaneously impose
\be
\g_1 \th^{12}=\th^{12} , ~~~
\g_1 \th^{34}=-\th^{34},
\ee
since $\th^{12}$ and $\th^{34}$ are not related. This means that $m_1=m_2=-m_3=-m_4=i$, and
\be
\th^{13}=\th^{14}=\th^{23}=\th^{24}=0.
\ee
Using (\ref{z4}), we have
\bea
&& \bar\th_{12}\g_1=\bar\th_{12}, ~~~
   \bar\th_{34}\g_1=-\bar\th_{34},  \nn\\
&& \bar\th_{13}=\bar\th_{14}=\bar\th_{23}=\bar\th_{24}=0.
\eea
It is similar for the conformal SUSY transformation parameters $\vth^{IJ}$ and $\bar\vth_{IJ}$.
So there are 1/6 BPS Wilson loops along infinite straight lines, as well as circular 1/6 BPS Wilson loops in Euclidean ABJM theory. They were considered in \cite{Drukker:2008zx,Chen:2008bp,Rey:2008bh}.

Then we consider the Wilson loop (\ref{w12abjm}) in Euclidean space along the line $x^\m=\t \d^\m_1$. We suppose
\be
\g_1\th^{1i}=\th^{1i}, ~~~ \g_1\th^{ij}=-\th^{ij},
\ee
and from (\ref{z4}) we have
\be
\bar \th_{1i}\g_1=\bar\th_{1i}, ~~~ \bar \th_{ij}\g_1=-\bar\th_{ij}.
\ee
We choose $ m_1=n_1=i$ and $ m_2=n_2=-i$, and for Poincar\'e SUSY transformation we get
\bea
&& \d_\th\mA=-\f{4\pi}{k} \lt( \phi_i\bar\psi_1\th^{1i}+\bar\th_{1i}\psi^1\bar\phi^i \rt), \nn\\
&& \d_\th\hat\mA=-\f{4\pi}{k} \lt(\bar\psi_1\th^{1i}\phi_i+\bar\phi^i\bar\th_{1i}\psi^1 \rt).
\eea
We have to choose
\be
\g_1 \eta=\eta, ~~~ \bar\zeta \g_1=\bar\zeta,
\ee
and we get
\bea
&& \d_\th \bar f_1=\sr{\f{2\pi}{k}}\bar\zeta\th^{1i} \mD_\t\phi_i, ~~~
   \bar g_1=\sr{\f{2\pi}{k}}\bar\zeta\th^{1i} \phi_i, \nn\\
&& \d_\th f_2=-\sr{\f{2\pi}{k}}\bar\th_{1i}\eta \mD_\t\bar\phi^i, ~~~
   g_2=-\sr{\f{2\pi}{k}}\bar\th_{1i}\eta \bar\phi^i .
\eea
Then, given
\be
\eta \bar\zeta=i(1+\g_1),
\ee
1/2 Poincar\'e SUSY is preserved. It is similar for conformal SUSY transformation.
So there is 1/2 BPS Wilson loop along an infinite straight line.

Also there is circular 1/2 BPS Wilson loop along $x^\m=(\cos\t,\sin\t,0)$.
The preserved SUSY is
\be
\vth^{1i}=i\g_3\th^{1i}, ~~~ \vth^{ij}=-i\g_3\th^{ij},
\ee
with $i,j=2,3,4$. The circular 1/2 BPS Wilson loops have been studied in \cite{Drukker:2009hy}.

\section{General curves in Minkowski spacetime}\label{general}

There can be BPS Wilson loops of general curves other than straight lines and circles \cite{Zarembo:2002an,Drukker:2007dw, Drukker:2007qr,Griguolo:2012iq,Cardinali:2012ru,Kim:2013oza,Bianchi:2014laa,Correa:2014aga}. In this section we show that in Minkowski spacetime BPS Wilson loops along general curves are necessarily timelike or null.

\subsection{\emph{d}=4 $\bf\mN$=4 SYM theory}\label{r1}

Along a general curve $x^\m(\t)$ we define the Wilson loop%
\footnote{Note that the definition of $y^I$ here is not the same as that in (\ref{wlde4ne4}). In the present case we have absorbed the factor $|\dot x|$ in $y^I$ for the convenience of subsequent discussions. Similar definitions would happen below for Wilson loops (\ref{hh1}), (\ref{e6}), and (\ref{hh2}).}
\be
W=\mP \exp \lt[ {ig\int d\t \lt( A_\m \dot x^\m+\phi_I y^I \rt)} \rt],
\ee
with $y^I$ being a function of $\t$ and transforming in reparameterization as
\be
y^I(\t)=y^I(\t')\f{\dd \t'}{d\t}.
\ee
We want the Wilson loop to preserve at least one supersymmetry that is parameterized by a ten-dimensional constant Weyl-Majorana spinor $\e$, and we need
\be \label{e3}
( \dot x^\m \g_\m+\g_I y^I)\e=0.
\ee
Taking complex conjugate of the above equation, we have
\be
(\g_\m^* \dot x^\m+\g_I^* y^{I*})\e^*=0.
\ee
Since $\e$ is a Majorana spinor, we get
\be \label{e4}
( \dot x^\m \g_\m+\g_I y^{I*})\e=0.
\ee
Using (\ref{e3}) and (\ref{e4}), we have $(y^I-y^{I*})\g_I \e=0$. Then we get
\be
0=[(y^I-y^{I*})\g_I]^2\e=(y^I-y^{I*})(y_I-y_I^*)\e,
\ee
Since $\e \neq 0$, we see that $y^I$ is real
\be y^{I*}=y^I. \ee
From (\ref{e3}) we get
\be
0=( \dot x^\m \g_\m+\g_I y^I)^2\e=(\dot x_\m\dot x^\m+ y^I y_I)\e,
\ee
which means that
\be
\dot x_\m\dot x^\m=- y^I y_I \leq 0
\ee
When $y^I=0$ for all $I$, we obtain $\dot x_\m\dot x^\m=0$ and the curve is null. When $y^I \neq 0$, we have $\dot x_\m\dot x^\m<0$ and the curve is timelike. So a BPS Wilson loop in $d=4$ $\mN=4$ SYM theory in Minkowski spacetime must be timelike or null.

\subsection{\emph{d}=3 $\bf\mN$=2 SCSM theory}\label{r2}

Along a general curve $x^\m(\t)$ we define the Wilson loop
\bea \label{hh1}
&& W=\mP \exp\lt( -i\int d\t\mA(\t) \rt) ,\nn\\
&& \mA=A_\m \dot x^\m+\f{2\pi}{k} m\phi\bar\phi,
\eea
with $m$ being a function of $\t$. To make the Wilson loop SUSY invariant we need $\d_\chi \mA=0$ for some nonvanishing $\chi$ and $\bar\chi$, and we have
\be
 \dot x^\m \g_\m \chi=i m \chi, ~~~ \bar\chi \dot x^\m \g_\m=i m \bar\chi.
\ee
Taking the complex conjugate of the second equation we have
\be
 \dot x^\m \g_\m \chi=i m^* \chi.
\ee
Then we get $(m-m^*)\chi=0$. Since $\chi \neq0$, we have $m=m^*$. We also have
\be
\dot x^\m \dot x_\m\chi=( \dot x^\m \g_\m)^2 \chi=-m^2 \chi,
\ee
which means
\be
\dot x^\m \dot x_\m=-m^2 \leq 0.
\ee
When $m \neq 0$, the BPS Wilson loop is timelike. When $m = 0$, the BPS Wilson loop is null. But it cannot be spacelike.

\subsection{ABJM theory}\label{s4.3}

For the ABJM theory in Minkowski spacetime we consider general GY type and DT type BPS Wilson loops.

\subsubsection{GY type Wilson loop}

We consider the Wilson loop along a general curve $x^\m(\t)$
\bea \label{e6}
&& W=\mP \exp \lt( -i\int d\t\mA(\t) \rt),\nn\\
&& \mA=A_\m \dot x^\m +\f{2\pi}{k} M^I_{\ph{I}J} \phi_I\bar\phi^J,
\eea
with $M^I_{\ph{I}J}$ being a $4\times4$ complex matrix and dependent on $\t$. To make the Wilson loop SUSY invariant we need
\be \label{e5}
 \dot x^\m \g_\m \chi^{IJ}=-i M^I_{\ph{I}K} \chi^{KJ}, ~~~
\bar \chi_{IJ} \dot x^\m \g_\m=-i M^K_{\ph{K}I} \bar\chi_{KJ},
\ee
with at least one component of $\chi^{IJ}$ being nonvanishing. Taking complex conjugate of the second equation we have
\be
 \dot x^\m \g_\m \chi^{IJ}=-i M^{\dagger I}_{\ph{\dagger I}K} \chi^{KJ},
\ee
with the matrix $M^\dagger$ being the Hermitian conjugate of $M$
\be
M^{\dagger I}_{\ph{\dagger I}J}=(M^J_{\ph{J}I})^*.
\ee
Then we have
\be
\dot x^\m \dot x_\m \chi^{IJ}= ( \dot x^\m \g_\m)^2 \chi^{IJ} = -A^I_{\ph{I}K} \chi^{KJ},
\ee
with $A$ being a positive semi-definite Hermitian matrix
\be
A^I_{\ph{I}J}=M^{\dagger I}_{\ph{\dagger I}K} M^K_{\ph{K}J},
\ee
whose eigenvalues can only be real positive or vanishing.
We have at least one $J=J_0$ that makes $\chi^{IJ_0}\neq0 $. Then
\be
A^I_{\ph{I}K} \chi^{KJ_0}= -\dot x^\m \dot x_\m \chi^{IJ_0}.
R\ee
It is just the eigenvalue equation of $A$, and $-\dot x^\m \dot x_\m$ is one eigenvalue. Then we have
\be
\dot x^\m \dot x_\m \leq 0.
\ee
Thus the BPS GY type Wilson loop can only be timelike or null.

\subsubsection{DT type Wilson loop}

We consider the DT type Wilson loop along a general curve $x^\m(\t)$
\bea \label{hh2}
&& W=\mP \exp \lt( -i\int d\t L(\t) \rt),                                  ~~~
   L=\lt( \ba{cc} \mA &\bar f_1 \\ f_2 & \hat\mA \ea \rt),                 \nn\\
&& \mA=A_\m\dot x^\m+\f{2\pi}{k} M^I_{\ph IJ} \phi_I\bar\phi^J,            ~~~
   \hat\mA=\hat A_\m\dot x^\m+\f{2\pi}{k} N_I^{\ph IJ}\bar\phi^I \phi_J,   \\
&& \bar f_1=\sr{\f{2\pi}{k}}\bar\zeta_I\psi^I,                              ~~~
   f_2=\sr{\f{2\pi}{k}}\bar\psi_I\eta^I,                                   \nn
\eea
with $M^I_{\ph IJ}$, $N_I^{\ph IJ}$, $\bar\zeta_I$ and $\eta^I$ being functions of $\t$.
In literature, all the DT type Wilson loops that were investigated in \cite{Drukker:2009hy,Griguolo:2012iq,Cardinali:2012ru,Kim:2013oza,Bianchi:2014laa,Correa:2014aga} belong to the class of Wilson loops that have at least local $SU(3)$ R-symmetry, since this is required if the Wilson loop has simple fundamental string worldsheet dual. However, we make no such assumption here, and investigate the general case.

In order to make the Wilson loop BPS we need to find $\bar g_1$ and $g_2$ that satisfy (\ref{e2}). One of the consequences is that
\be
\bar g_1= \sr{\f{2\pi}{k}} \bar\a^I \phi_I, ~~~ g_2= -\sr{\f{2\pi}{k}} \b_I \bar\phi^I,
\ee
with $\bar \a^I$ and $\b_{I}$ being Grassmann odd and having no free color index or spinor index. We also have
\bea
&&  \dot x^\m \g_\m \chi^{IJ}=-i M^I_{\ph{I}K} \chi^{KJ} + i \bar\a^I \eta^J,                      \nn\\
&& \bar \chi_{IJ} \dot x^\m \g_\m=-i M^K_{\ph{K}I} \bar\chi_{KJ} + i \bar\zeta_J \b_I.
\eea
Taking the complex conjugate of the second equation we get
\be
 \dot x^\m \g_\m \chi^{IJ}=-i M^{\dagger I}_{\ph{\dagger I}K} \chi^{KJ} + i \bar\b^I \z^J  ,
\ee
with $\z^J= \bar\zeta_J^*$ and $\bar\b^I=\b_I^*$.

We consider an arbitrary fixed point on the curve, say the point $\t=\t_0$.  For the Wilson loop to be supersymmetric, we should have $\chi^{IJ}_\a \neq 0$ for some $I,J,\a$. Let us first consider the case with some $\chi^{IJ}_+$ being nonzero.%
\footnote{Notice that this condition is not affected by the  $SU(4)$ R-symmetry transformation.}
From \be \bar\chi_{IJ}=\f{1}{2}\e_{IJKL}\chi^{KL},  \ee
we know that there should be at least one $I$, such that $\chi^{I4}_+\neq 0$.
Then we perform an
 $SU(4)$ R-symmetry transformation such that
\be
\eta^4_+=\eta^4_-=\zeta^4_+=0.
\ee
Then we have
\bea
&& \dot x^\m \g_\m \chi^{I4}=-i M^{I}_{\phantom{I}J} \chi^{J4}, \nn\\
&& \dot x^\m \g_{\m+}^{\phantom{\m+}\b} \chi^{I4}_\b=-i M^{\dagger I}_{\phantom{\dagger I}J} \chi^{J4}_+.
\eea
Note the first equation applies to both spinor indices, but the second one only applies to index $+$.
Then we have
\be
\dot x_\m\dot x^\m \chi^{I4}_+  = \dot x^\m \dot x^\n \g_{\m+}^{\phantom{\m+}\a} \g_{\n\a}^{\phantom{\m+}\b} \chi^{I4}_\b
                      = -M^{I}_{\phantom{I}J} M^{\dagger J}_{\phantom{\dagger I}K} \chi^{K4}_+.
\ee
Similar to the discussion of the previous subsection, we see that the curve at $\t=\t_0$ must be timelike or null
\be
\dot x_\m\dot x^\m \leq 0.
\ee

Now we are left with the case  $\chi^{IJ}_+ = 0$ for all $I,J$. Then for the Wilson loop to be BPS, we should have $\chi^{IJ}_- \neq 0$ for some $I,J$, which is equivalent to the statement that $\chi^{I4}_- \neq 0$ for some $I$. In this case we perform an $SU(4)$ R-symmetry transformation such that
\be
\eta^4_+=\eta^4_-=\zeta^4_-=0,
\ee
from which we get
\bea
&& \dot x^\m \g_\m \chi^{I4}=-i M^{I}_{\phantom{I}J} \chi^{J4}, \nn\\
&& \dot x^\m \g_{\m-}^{\phantom{\m-}\b} \chi^{I4}_\b=-i M^{\dagger I}_{\phantom{\dagger I}J} \chi^{J4}_-.
\eea
Then we have
\be
\dot x_\m\dot x^\m \chi^{I4}_-= -M^{I}_{\phantom{I}J} M^{\dagger J}_{\phantom{\dagger I}K} \chi^{K4}_-.
\ee
We still have that the curve at $\t=\t_0$ must be timelike or null
\be
\dot x_\m\dot x^\m \leq 0.
\ee

So we always have $\dot x^\m \dot x_\m \leq 0$ at point $\t=\t_0$. Since the point is chosen arbitrarily, we have $\dot x^\m \dot x_\m \leq 0$ everywhere on the curve. Thus the DT type BPS Wilson loop in ABJM theory in Minkowski spacetime must be timelike or null.

\section{Conclusion and discussion}\label{s4}

We have discussed BPS Wilson loops in several superconformal theories, namely the $d=4$ $\mN=4$ SYM theory, the $d=3$ $\mN=2$ SCSM theory, and the ABJM theory. We found that in Minkowski spacetime there exist BPS Wilson loops along timelike and null infinite straight lines, but there are no BPS Wilson loops along spacelike infinite straight lines or circles. However, in Euclidean space BPS Wilson loops are allowed for both spacelike infinite straight lines and circles. Furthermore, we give general proofs that BPS Wilson loops in these superconformal theories in Minkowski space must be timelike or null.

The result is plausible in view of AdS/CFT correspondence. A one-dimensional BPS Wilson loop in a superconformal field theory is dual to the two-dimensional worldsheet of the fundamental string in AdS space. The extra spacelike dimension is just along the AdS radial direction. If the BPS Wilson loop is timelike or null, then the string worldsheet is also timelike or null. This certainly can only happen in Minkowski spacetime. If the BPS Wilson loop is spacelike, then the string worldsheet is also spacelike. This can only happen in Euclidean space.
A spacelike brane, or an S-brane, in Minkowski spacetime cannot preserve any SUSY\cite{Gutperle:2002ai,Sen:2002nu,Sen:2002in,Sen:2002an}.
This explains why there is no BPS spacelike Wilson loop in Minkowski spacetime.

\section*{Acknowledgments}

We would like to thank Bin Chen, Jian-Xin Lu and Zohar Komargodski for valuable discussions.
Special thanks to the anonymous referee for valuable suggestions that there should be proofs that BPS Wilson loops along general curves in Minkowski spacetime cannot be spacelike. The referee also provided us with the calculation details of Subsection~\ref{r1} and \ref{r2}, and gave us valuable suggestions about the DT type Wilson loops in ABJM theory. Without the criticism and insistence of the referee we could not have completed the work.
JW would like to thank KIAS and ICTS-USTC for hospitality during recent visits.
The work was in part supported by NSFC Grants No.~11105154, No.~11222549 and No.~11575202. JW also gratefully acknowledges the support of K.~C.~Wong Education Foundation and Youth Innovation Promotion Association of CAS  (No.~2011016).

\begin{appendix}

\section{Majorana spinors in various dimensions}\label{sa}

In this appendix we review the definitions of Majorana spinors in various dimensions. We follow closely the Appendix~B of \cite{Polchinski:1998rr}, and one can find details therein.

In $d$-dimensional Minkowski spacetime, the gamma matrices $\g_\m$ that form Clifford algebra
\be
\lt\{ \g_\m, \g_\n  \rt\}=2\eta_{\m\n}.
\ee
Here we use the mostly plus metric $\eta_{\m\n}=\diag(-,+,+,\cdots)$. Often, one requires
\be
\g_\m^\dagger=\g_0 \g_\m \g_0.
\ee
The matrices $\pm \g_\m^*$ also satisfy the Clifford algebra, and so there must be similarity transformation
\be  \label{alpha}
B\g_\m B^{-1}=\pm \g_\m^* \equiv (-)^\a \g_\m^*.
\ee
Here we use $*$ as complex conjugate, and define $\a=0$ for the plus sign and $\a=1$ for minus sign. Given a Dirac spinor $\th$ one can define the charge conjugate
\be \label{thc}
\th_c \equiv B^{-1}\th^*.
\ee
The spinor $\th_c$ transforms the same way as $\th$ under the Lorentz transformation. When $B$ satisfies
\be \label{B}
B^*=B^{-1},
\ee
we can impose the reality condition
\be \label{Majorana}
\th = \th_c,
\ee
and get a Majorana spinor.
We list the dimensions of Minkowski spacetime in which the Majorana spinors are allowed and the corresponding $\a$ as below.
\begin{center}
\begin{tabular}{c|c|c|c|c|c|c|c|c|c|c}
$d$  & \multicolumn{2}{c|}{2} & 3 & 4 & 8 & 9 & \multicolumn{2}{c|}{10}  & 11 & 12\\\hline
$\a$ & 0            & 1       & 0 & 0 & 1 & 1 &       0  & 1             & 0  & 0 \\
\end{tabular}
\end{center}

Under a general similarity transformation
\be
\td\g_\m=U \g_\m U^{-1},
\ee
we have
\be
\td\th=U\th, ~~~ \td B =U^* B U^{-1}.
\ee
To preserve
\be
\td\g_\m^\dagger=\td\g_0 \td\g_\m \td\g_0,
\ee
we need $U$ to be unitary
\be
U^\dag=U^{-1}.
\ee
One can show that $\a$ defined in (\ref{alpha}), the criterion that reality condition can be imposed (\ref{B}), and the definition of Majorana spinors (\ref{Majorana}) do not change under this similarity transformation.

\section{Consistent constraints for Majorana spinors}\label{sb}

Constraints on spinors are often used in physics, for example in search of BPS objects in supersymmetric theories. Sometimes the spinors are Majorana spinors. For a Majorana spinor there is already the reality condition as reviewed in  the previous appendix. Other constraints should be consistent with this reality condition. As an example, in even dimensions one can impose the chirality constraint for the Dirac spinors and get Weyl spinors.
In four-dimensional Minkowski spacetime, the chirality constraint of the Weyl spinor is not consistent with the reality condition of the Majorana spinor.
So although there are both Weyl and Majorana spinors, there are no Weyl-Majorana spinors in four-dimensional spacetime.
In this appendix we investigate the consistent constraints of Majorana spinors in dimensions $2 \leq d \leq 12$ when Majorana spinors exist.

We first consider $d$-dimensional Minkowski spacetime. When $d=2k+2$, there are linearly independent matrices
\be
\g_{\m_1 \cdots \m_n} \equiv  \g_{[\m_1}\cdots\g_{\m_n]}    , ~~~ n=1,2,\cdots,2k+2.
\ee
When there is $\g_0$ in $\g_{\m_1 \cdots \m_n}$, we say $\b=1$, otherwise we say $\b=0$. It is easy to show that
\be
\tr \g_{\m_1 \cdots \m_n}=0, ~~~ \lt( \g_{\m_1 \cdots \m_n} \rt)^2=(-)^{\b+\f{n(n-1)}{2}}.
\ee
Note that there is no summation of indices in the second equation. Sometimes we want to use matrix $\g_{\m_1 \cdots \m_n}$ to construct a constraint and eliminate half of the degree of freedom of a Majorana spinor $\th$ by the constraint equation%
\footnote{Equivalently, we may define the projection operator
\be P^\pm_{\m_1 \cdots \m_n}= \f{1}{2} \lt( 1 \pm (-i)^{\b+\f{n(n-1)}{2}} \g_{\m_1 \cdots \m_n} \rt). \nn\ee
The constraint equation (\ref{coneq}) is just the projection equation
\be P^+_{\m_1 \cdots \m_n}\th=\th, \nn \ee
or
\be P^-_{\m_1 \cdots \m_n}\th=0. \nn \ee}
\be \label{coneq}
\g_{\m_1 \cdots \m_n} \th =i^{\b+\f{n(n-1)}{2}} \th.
\ee
For a Dirac spinor, it is fine, but for a Majorana spinor there is subtlety.
We take complex conjugate of the equation, use the Majorana condition, and finally get
\be
\g_{\m_1 \cdots \m_n} \th =(-)^{n\a+\b+\f{n(n-1)}{2}}i^{\b+\f{n(n-1)}{2}} \th.
\ee
For $\th \neq 0$, we need
\be
(-)^{n\a+\b+\f{n(n-1)}{2}}=1.
\ee
The solutions are listed below.
\begin{center}
\begin{tabular}{c|c|c}
        & $\b=1$ & $\b=0$ \\\hline
 $\a=0$ & $n=2,3$ mod 4 & $n=0,1$ mod 4 \\\hline
 $\a=1$ & $n=1,2$ mod 4 & $n=0,3$ mod 4
\end{tabular}
\end{center}
When $d=2k+3$, there is the constraint
\be
\g_0 \g_1 \cdots \g_{2k+2}=\pm i^k, \ee
where the sign can be chosen arbitrarily.
Then the linearly independent matrices are
\be
\g_{\m_1 \cdots \m_n} \equiv \g_{[\m_1}\cdots\g_{\m_n]}    , ~~~ n=1,2,\cdots,k+1.
\ee
The condition for them to be consistent as constraint matrices for a Majorana spinor is the same as before.

\begin{table}[!hbp]
\centering
\begin{tabular}{c|c|c}\hline\hline
 $d$ & $\a$ & Consistent constraint matrices \\\hline
  \multirow{2}*{2}  & 0                 & $\g_{01}$, $\g_1$ \\ \cline{2-3}
                    & 1                 & $\g_0$, $\g_{01}$ \\ \hline
  3                 & 0                 & $\g_i$/$\g_{0i}$ \\ \hline
  4                 & 0                 & $\g_{0i}$, $\g_{0i_1i_2}$, $\g_i$  \\ \hline

  \multirow{2}*{8}  & \multirow{2}*{1}  & $\g_{0}$, $\g_{0i}$, $\g_{0i_1\cdots i_4}$, $\g_{0i_1\cdots i_5}$,  \\
                    &                   & $\g_{i_1\cdots i_3}$, $\g_{i_1\cdots i_4}$, $\g_{1\cdots 7}$ \\ \hline

  \multirow{2}*{9}  & \multirow{2}*{1}  & $\g_0$/$\g_{1\cdots 8}$, $\g_{0i}$/$\g_{i_1\cdots i_7}$,  \\
                    &                   & $\g_{i_1\cdots i_3}$/$\g_{0i_1\cdots i_5}$,
                                          $\g_{i_1\cdots i_4}$/$\g_{0i_1\cdots i_4}$  \\ \hline

                    & \multirow{2}*{0}  & $\g_{0i}$, $\g_{0i_1 i_2}$, $\g_{0i_1\cdots i_5}$, $\g_{0i_1\cdots i_6}$, $\g_{01\cdots 9}$, \\
  \multirow{2}*{10} &                   & $\g_i$, $\g_{i_1\cdots i_4}$, $\g_{i_1\cdots i_5}$,
                                          $\g_{i_1\cdots i_8}$, $\g_{1\cdots 9}$ \\ \cline{2-3}
                    & \multirow{2}*{1}  & $\g_{0}$, $\g_{0i}$, $\g_{0i_1\cdots i_4}$, $\g_{0i_1\cdots i_5}$,
                                          $\g_{0i_1\cdots i_8}$, $\g_{01\cdots 9}$, \\
                    &                   & $\g_{i_1\cdots i_3}$, $\g_{i_1\cdots i_4}$,
                                          $\g_{i_1\cdots i_7}$, $\g_{i_1\cdots i_8}$ \\ \hline

  \multirow{2}*{11} & \multirow{2}*{0}  &  $\g_{0i}$/$\g_{i_1\cdots i_{9}}$, $\g_{0i_1i_2}$/$\g_{i_1\cdots i_8}$,  \\
                    &                   &  $\g_i$/$\g_{0i_1\cdots i_{9}}$, $\g_{i_1\cdots i_4}$/$\g_{0i_1\cdots i_6}$,
                                           $\g_{i_1\cdots i_5}$/$\g_{0i_1\cdots i_5}$  \\ \hline

  \multirow{2}*{12} & \multirow{2}*{0}  & $\g_{0i}$, $\g_{0i_1 i_2}$, $\g_{0i_1\cdots i_5}$, $\g_{0i_1\cdots i_6}$,
                                          $\g_{0i_1\cdots i_9}$, $\g_{0i_1\cdots i_{10}}$, \\
                    &                   & $\g_i$, $\g_{i_1\cdots i_4}$, $\g_{i_1\cdots i_5}$,
                                          $\g_{i_1\cdots i_8}$, $\g_{i_1\cdots i_9}$ \\\hline\hline
\end{tabular}
\caption{Consistent constraint matrices for Majorana spinors in Minkowski spacetime. Here the Latin letters $i,i_1,i_2,\cdots$ vary from 1 to $d-1$. Matrices separated by ``/'' are just the equivalent ones, up to a possible factor $-1$ or $\pm i$.}\label{t1}
\end{table}

\begin{table}[!hbp]
\centering
\begin{tabular}{c|c|c}\hline\hline
 $d$ & $\a$ & Consistent constraint matrices \\ \hline

  2                 & 0                 & $\g_\m$ \\ \hline

  6                 & 1                 & $\g_{\m_1\cdots \m_3}$, $\g_{\m_1\cdots \m_4}$ \\ \hline

  7                 & 1                 & $\g_{\m_1\cdots \m_3}$/$\g_{\m_1\cdots \m_4}$ \\ \hline

  \multirow{2}*{8}  & 0                 & $\g_{\m}$, $\g_{\m_1\cdots \m_4}$, $\g_{\m_1\cdots \m_5}$
                                          $\g_{1\cdots 8}$ \\ \cline{2-3}
                    & 1                 & $\g_{\m_1\cdots \m_3}$, $\g_{\m_1\cdots \m_4}$,
                                          $\g_{\m_1\cdots \m_7}$, $\g_{1\cdots 8}$ \\\hline

  9                 & 0                 & $\g_{\m}$/$\g_{\m_1\cdots \m_8}$,
                                          $\g_{\m_1\cdots \m_4}$/$\g_{\m_1\cdots \m_5}$  \\ \hline

  10                & 0                 &  $\g_{\m}$, $\g_{\m_1\cdots \m_4}$, $\g_{\m_1\cdots \m_5}$,
                                           $\g_{\m_1\cdots \m_8}$, $\g_{\m_1\cdots \m_9}$ \\\hline\hline
\end{tabular}
\caption{Consistent constraint matrices for Majorana spinors in Euclidean space. Here the Greek letters $\m,\m_1,\m_2,\cdots$ vary from 1 to $d$. }\label{t2}
\end{table}

In summary, we list all the possible consistent constraint matrices of Majorana spinors as in Table~\ref{t1}.
Note that when $d=2$ there is matrix $\g_{01}$ and when $d=10$ there is matrix $\g_{01\cdots9}$, which is just that there are Weyl-Majorana spinors in these dimensions. For Weyl-Majorana spinors, a constraint matrix must has even number of gamma matrices. In two dimensions, there is no constraint matrix for Weyl-Majorana spinors. In ten dimensions, the consistent constraint matrices are
\be
\g_{0i}, \g_{0i_1\cdots i_5}, \g_{i_1\cdots i_4}, \g_{i_1\cdots i_8}.
\ee

For $d$-dimensional Euclidean space the metric is $\d_{\m\n}=\diag(+++\cdots)$, and the Clifford algebra becomes
\be
\lt\{ \g_\m, \g_\n  \rt\}=2\d_{\m\n}.
\ee
The analysis method is the same as before, and we investigate $2 \leq  d \leq 12$ when Majorana spinors exist. The final results are listed as in Table~\ref{t2}.
When $d=8$ there is constraint matrix $\g_{1\cdots 8}$ and this just means the existence of Weyl-Majorana spinors. Now the consistent constraint matrices of the Weyl-Majorana spinors are
\be
\g_{\m_1\cdots \m_4}.
\ee

As an application of the above discussions we revisit one problem in Subsection~3.4 of \cite{Drukker:2008zx}.
There were analyses of the Killing spinors of $d=11$ M-theory in the AdS$_4\times$S$^7$/Z$_k$ spacetime. One needs
\be \label{dpy}
( \g_{47}+\g_{58}+\g_{69}+\g_{456789} )\e_0=0,
\ee
and $\e_0$ is a constant Majorana spinor. The authors used $\g_{47}$, $\g_{58}$, $\g_{69}$, and $\g_{456789}$ as constraint matrices, but from above discussions this is illegal. We rewrite (\ref{dpy}) as
\be \label{wz}
( \g_{4578}+\g_{4679}+\g_{5689}+1 )\e_0=0.
\ee
Because $[\g_{4578},\g_{4679}]=0$, we can use the basis in which
\be
\g_{4578}\e_0=s_1 \e_0, ~~~ \g_{4679}\e_0=s_2 \e_0,
\ee
and so we get
\be
\g_{5689}\e_0=s_1 s_2 \e_0.
\ee
Here $s_{1,2}$ are $\pm1$, and so the Majorana spinor $\e_0$ takes four configurations
\be
(s_1,s_2)=(++), (+-), (-+), (--).
\ee
Among them only the first one does not satisfy (\ref{wz}). So 1/4 supercharges are broken in the orbifolding. We can also proceed with this and analyze the supercharges preserved by fundamental string (without or with smearing), D2-brane (without or with smearing), and D6-brane in the orbifold spacetime. The process is similar to what is discussed above and the final conclusions in \cite{Drukker:2008zx} do not change.

\section{Conventions in \emph{d}=3 Minkowski spacetime}\label{sc}

We follow most of the conventions in \cite{Benna:2008zy}, but there are also some minor differences. In three-dimensional Minkowski spacetime, we use the coordinates $x^\m=(x^0,x^1,x^2)$ and the metric $\eta_{\m\n}=\diag(-++)$. We choose the gamma matrices as
\be
\g^{\m\phantom{\a}\b}_{\phantom{\m}\a}=(i\s^2,\s^1,\s^3),
\ee
with $\s^{1,2,3}$ being the Pauli matrices. Note that these are real matrices. They satisfy $\g^\m\g^\n=\eta^{\m\n}+\e^{\m\n\r}\g_\r$, $\g_\m\g_\n=\eta_{\m\n}+\e_{\m\n\r}\g^\r$, with $\e^{\m\n\r}$ and $\e_{\m\n\r}$ being totally antisymmetric and $\e^{012}=-\e_{012}=1$.

We have the Grassmann odd spinor $\th_\a$ with the spinor index $\a=+,-$. We define the matrices
\be \label{epsilon}
\e^{\a\b}=\lt( \ba{cc} & 1 \\ -1 & \ea \rt), ~~~ \e_{\a\b}=\lt( \ba{cc} & -1 \\ 1 & \ea \rt).
\ee
Spinor indices are raised and lowered as
\be
X^\a=\e^{\a\b}X_\b, ~~~ X_\a=\e_{\a\b}X^\b.
\ee
Then one can get
\be
Y^\a_{\ph \a\a}=-Y_\a^{\ph\a\a}.
\ee
Here $X$ and $Y$ are general objects with spin indices, but they cannot involve $\e^{\a\b}$, $\e_{\a\b}$, $\d_\a^\b$, $\p_\a$, or $\p^\a$. Thus we have
\be
\g^\m_{\a\b} \equiv \e_{\b\g}\g^{\m\ph\a\g}_{\ph\m\a}= \lt( -1, -\s^3, \s^1 \rt),
\ee
which are real symmetric matrices. The conventions allow us to define the charge conjugate of spinors as
\be
\bar\th_\a = \th^*_\a, ~~~ \bar\th_\a^*=\th_\a.
\ee
It is easy to see $\bar{\bar\th}=\th$. We also define the shorthand
\be
\th\psi\equiv\th^\a\psi_\a, ~~~ \th\g^\m\psi \equiv \th^\a\g^{\m\ph\a\b}_{\ph\m\a} \psi_\b .
\ee
We have the following useful relations
\bea
&& \th\psi=\psi\th, ~~~  (\th\psi)^*=-\bar\psi\bar\th, ~~~ \g^\m\th=-\th\g^\m, \nn\\
&&\th\g^\m\psi=-\psi\g^\m\th, ~~~ (\th\g^\m\psi)^*=\bar\psi\g^\m\bar\th.
\eea

\section{Conventions in \emph{d}=3 Euclidean space}\label{sd}

In three-dimensional Euclidean spacetime, we use the coordinates $x^\m=(x^1,x^2,x^3)$ and the metric $\d_{\m\n}=\diag(+++)$. We choose the gamma matrices as
\be
\g^{\m\phantom{\a}\b}_{\phantom{\m}\a}=(-\s^2,\s^1,\s^3),
\ee
with $\s^{1,2,3}$ being the Pauli matrices.
Note that $\lt(\g^\m \rt)^\dagger=\g^\m$, i.e.
$\lt( \g^{\m\phantom{\a}\b}_{\phantom{\m}\a} \rt)^*=\g^{\m\phantom{\b}\a}_{\phantom{\m}\b}$.
 We have $\g^\m\g^\n=\d^{\m\n}+i\e^{\m\n\r}\g_\r$, $\g_\m\g_\n=\d_{\m\n}+i\e_{\m\n\r}\g^\r$, with $\e^{\m\n\r}$ and $\e_{\m\n\r}$ being totally antisymmetric and $\e^{123}=\e_{123}=1$.

We have the spinor $\th_\a$ that is Grassmann odd. The spinor indices $\a,\b,\cdots$ can be raised or lowered using $\e^{\a\b}$ or $\e_{\a\b}$ in the same way as the Minkowski case. One can check that $\g^\m_{\a\b}$ is symmetric but not real. Note that there is no Majorana spinor in $d=3$ Euclidean space. From $\th$ there can be spinor $\th^\dagger$ satisfying
\bea
&& \th_\a^*=\th^{\dagger\a}, ~~~ \th^{\a*}=-\th^\dagger_\a, \nn\\
&& \th^{\dagger\a*}=\th_\a, ~~~ \th^{\dagger*}_\a=-\th^{\a}.
\eea
Formally we have $\th^{\dagger\dagger}=-\th$. However, $\th^\dagger$ will not be used in this paper.
We also have symbol $\bar\th$, but it is independent and has nothing to do with $\th$, $\th^*$ or $\th^\dagger$.
There are shorthand the same as the Minkowski case
\be
\th\psi\equiv\th^\a\psi_\a, ~~~ \th\g^\m\psi \equiv \th^\a\g^{\m\ph\a\b}_{\ph\m\a} \psi_\b .
\ee
We have the following relations
\bea
&& \th\psi=\psi\th, ~~~  (\th\psi)^*=-\psi^\dagger\th^\dagger, ~~~ \g^\m\th=-\th\g^\m, \nn\\
&&\th\g^\m\psi=-\psi\g^\m\th, ~~~ (\th\g^\m\psi)^*=-\psi^\dagger\g^\m\th^\dagger.
\eea

\end{appendix}



\providecommand{\href}[2]{#2}\begingroup\raggedright\endgroup

\end{document}